\documentstyle[12pt]{article}
\begin{document}

\title
{\Large \bf 
Extracting Messages Masked by Chaotic Signals of Time-delay Systems}

\author{ Changsong Zhou$^1$ and  C.-H. Lai$^{1,2}$ \\
        $^1$Department of Computational Science\\
        and $^2$Department of Physics\\
        National University of Singapore,
        Singapore 119260}

\date{}
\maketitle
\baselineskip 12pt

\begin{center}
\begin{minipage}{14cm}

\centerline{\bf Abstract}
\bigskip
\baselineskip 12pt

We show how to extract messages masked by a chaotic signal of
a time-delay system with very high dimension and many positive
Lyapunov exponents. Using a special embedding coordinate, the infinite
dimensional phase space of the time-delay system is projected to a special 
three-dimensional space, which enables us to identify  
the time-delay of the system from the transmitted signal, 
and reconstruct  the chaotic dynamics to unmask the hidden message 
successfully. 
The message extraction procedure is illustrated by simulations with Mackey-Glass
time-delay system for two type of masking schemes and different kinds of
messages.

PACS number(s): 05.45.+b;

\end{minipage}
\end{center}

\newpage

Application of chaotic synchronization systems to secure communication has
been a field of great research interest~\cite{pc,co, khec,pckh,kp}. 
However,  it has been shown that in some
low-dimensional chaotic systems with only one positive Lyapunov exponent,
the hidden message can be unmasked by dynamical reconstruction of the 
chaotic signal using nonlinear dynamical forecasting (NLDF)
 methods~\cite{short1, short2},
 by  some simple return maps~\cite{pc1}, or  some other methods~\cite{zhou}.
It has been suggested that one possible  way to improve the security is to employ hyperchaos 
in communication~\cite{kp, pdy, kps},
based on the consideration that increased randomness and unpredictability of the
hyperchaotic signals will make it more difficult to extract a masked message.
Lately, it has been shown that messages masked by hyperchaos of a
six-dimensional systems can also be attacked using the NLDF
methods~\cite{short3}, showing  that going to higer dimensions does not
produce a drastic improvement in the security of the system if the
local dynamics are still quite low dimensional.

It  has been  known  that very simple time-delay systems~\cite{mg}  are able
to  exhibit hyperchaos~\cite{farmer}.  Therefore,  it has been proposed in a recent 
report that time-delay systems provide  alternative simple and efficient tools for 
chaos communication with low detectability~\cite{ml}.
Chaotic attractors of time-delay systems can have much higher
dimension and many more positive Lyapunov exponents than the system studied in
Ref.~\cite{short3}, and whether 
the communication is as secure as expected  has not been examined yet. 
In this letter, we will show that messages masked by chaos of a time-delay
system with very high dimension  and many positive Lyapunov
exponents can be extracted successfully, not using some well-established
dynamical reconstruction  methods~\cite{short1, short2, short3}, but using 
a special yet simple embedding approach proposed recently for time-series
analysis of  time-delay systems~\cite{bpm1,bpm2,bpm3}.  

We focus our attention in this letter on scalar time-delay systems of
the form
\begin{equation}
\dot{x}=f(x, x_{\tau_0}),\;\;\;\;\;\; x_{\tau_0}=x(t-\tau_0). \label{eq_sys} 
\end{equation}
For such systems with large time-delay $\tau_0$, some well-established
nonlinear time-series analysis methods~\cite{gp,er,kba} run into severe 
problems~\cite{mp,hko}. 

An observation of the Eq.~(\ref{eq_sys}) shows that in a special 
three-dimensional space $(x_{\tau_0}, x, \dot{x})$, the dynamics of the 
system is restricted to a smooth  manifold defined by Eq.~(\ref{eq_sys}), namely
\begin{equation}
\dot{x}-f(x, x_{\tau_0})=0. 
\end{equation}
However, in a similar space $(x_{\tau}, x, \dot{x})$ with $\tau\neq \tau_0$,
the trajectory is no longer restricted to a smooth hypersurface, but fills
a great part of the space, resulting a more complicated structure. This makes it 
possible to detect the time-delay $\tau_0$ of the system by some measures of
the complexity as a function of embedding delay $\tau$, and then recover 
the dynamics of the system~\cite{bpm1,bpm2,bpm3}. 

This approach   is applied in this letter  
to extract messages masked by chaotic signals
of the above time-delay system. In the context of synchronization,
we consider the following communication system with two masking schemes 
considered in Ref.~\cite{ml}:

Scheme I:
\begin{eqnarray}
\dot{x}&=&f(x, x_{\tau_0})+kI, \nonumber \\
      s&=&x+I, \label{sch_i}
\end{eqnarray}
and Scheme II:
\begin{eqnarray}
\dot{x}&=&f(x, x_{\tau_0})+kIx, \nonumber\\
      s&=&x(1+I). \label{sch_ii}
\end{eqnarray}
An authorized receiver has an identical copy of the time-delay system
which is made synchronized with $x$ by the following coupling: 
\begin{equation}
\dot{y}=f(y, y_{\tau_0})+k(s-y).
\end{equation}
In this communication system, the message $I$,  
often much lower in amplitude  than the chaotic signal $x$, 
is injected into the transmitter to modulate the time-delay system. The 
injection of the message has effectively altered the transmitter dynamics and 
has been considered  as a way to improve the security~\cite{kp,ml}  compared  
with methods where the message is directly  added  to  
the chaotic carrier~\cite{co}. The masking Scheme II is expected to produce 
securer masking because the message and the chaotic signal couple with 
each other  in a more sophisticated manner. $s$ is the  signal transmitted to 
the receiver to achieve synchronization with a proper coupling parameter $k$.
As a third party, we do not  have a receiver system $y$, but have the
time series of the transmitted signal $s$ sampled by a time interval $h$.

Our message extraction approach consists of the following steps: 

(1) We project the time series $\{s^i\}$ to the three-dimensional space
$(s^i_{\tau}, s^i, \dot{s}^i)$ with $\dot{s}^i$ estimated as
 $\dot{s}^i= (s^{i+1}-s^{i-1})/2h$.

(2) We investigate the complexity of the projected trajectory  in the  
 $(s^i_{\tau}, s^i, \dot{s}^i)$ space by the measure of the smoothness.
First, we  apply a local linear approximation 
\begin{equation}
\hat{\dot{s}}=a_i+b_is+c_is_{\tau}
\end{equation}
to  a small  neighborhood $U_i$ of a point $(s^i_\tau, s^i)$.
The fitting parameters $a_i$, $b_i$ and $c_i$ are computed by least sequre fit, and 
the local  fitting error is
\begin{equation}
e_i=\frac{1}{M_{U_i}}\sum\limits_{j\in U_i}(\dot{s}^j-a_i-b_is^j-c_is^j_{\tau})^2,
\end{equation}
where $M_{U_i}$ is the number of the neighbor points.  
The average $E$ of $e_i$ over a number of points $(x^i_{\tau}, x^i$) provides 
a measure of the smoothness of the structure in the projected space. If
$\tau=\tau_0$, the trajectory is restricted to  close vicinity of  the smooth
 hypersurface for small enough message $I$, and $E$
can be rather small if the size $\epsilon$  of neighborhood is sufficiently small;
otherwise, $E$ can be quite larger because there is no unique functional
relationship between $\dot{s}$ and $(s_\tau, s)$ for $\tau\neq \tau_0$. 
We can expect a minimum of $E$ at $\tau=\tau_0$.
 By examining $E$ as a function of embedding delay $\tau$, we can detect 
the time-delay $\tau_0$ of the system by the minimum of $E$.   
  
(3) After correct identification of the value of $\tau_0$, we use the 
local linear approximation
\begin{equation}
\hat{\dot{x}}^i=a_i+b_is^i+c_is^i_{\tau_0}
\end{equation}
as an estimation of $\dot{x}=f(x, x_{\tau_0})$ of the time-delay system
 in the absence of message $I$. 
From Eqs~(3,4) we have 
$\dot{s}=f(x, x_{\tau_0})+kI+\dot{I}$  
for the  masking Scheme I, and
$\dot{s}=f(x, x_{\tau_0})+kIx+\dot{x}I+x\dot{I}$ for Scheme  II. 
For the  conditions  $|I|\ll |x|$, $|\dot{I}|\ll |kI|$, and $|\dot{x}|\ll |kx|$
with $| \cdot |$  denoting the  amplitude, 
 the extracted message can be estimated as 
\begin{equation}
kI^i_e=
\left\{
\begin{array}{ll}
\dot{s}^i-\hat{\dot{x}}^i, & \hbox{Scheme I},\\
(\dot{s}^i-\hat{\dot{x}}^i)/s^i, & \hbox{Scheme II}.
\end{array}
\right.
\end{equation}

To illustrate the message extraction procedure, we employ the
Mackey-Glass equation as in Ref.~\cite{ml},  
\begin{equation}
\dot{x}=f(x, x_{\tau_0})=-bx+\frac{ax_{\tau_0}}{1+x^c_{\tau_0}}.
\end{equation}
With parameter $b=0.1$,  $a=0.2$, and $c=10$,  the system is chaotic 
for $\tau_0>16.8$. In the  chaotic regime, the  number of positive Lyapunov 
exponents increases with $\tau_0$, and is about 15 for $\tau_0=300$, and the chaotic
attractor dimension increases almost linearly with $\tau_0$,  for  
example, the Kaplan-Yorke dimension is roughly 30 for $\tau_0=300$~\cite{ml}.
In our simulation, we take $\tau_0=300$ and $k=1.0$~\cite{ml}. 

In all the following examples, we record $N=50000$ points with sample interval
$h=0.5$. The size of neighborhood is set by $\epsilon=0.01$.   

First, let us consider a simple   message signal of  
sine wave  $I(t)=A\sin (2\pi t/T)$ with $A=0.005$ and $T=200$.
Fig. 1 shows the measure of smoothness $E$ as a function of $\tau$.
A pronounced minimum at $\tau=300$ enables  us to identify  the time-delay
of the system correctly although the system is modulated by the injected message $I$.
This is also true for our  other examples in the following where the
results of $E$ are not presented to save space. 

With the correct  value of time-delay, the message can be extracted
successfully, as illustrated in Fig. 2 for the masking Scheme I and Fig. 3
for the masking Scheme II, respectively. A comparison between the time
series of $s$ in Fig. 3(a) and that of $\Delta s=\dot{s}-\hat{\dot{x}}$   
in Fig. 3(b) reveals that when $s$ is close to zero in a certain 
 period of time, the corresponding $\Delta s$ is also close to zero in this
period of time,
indicating that $\Delta s$ is modulated by $s$. The demodulated signal 
$\Delta s/s$ is shown in Fig. 3 as the extracted message.
The results show that the masking Scheme II,  although can result in larger
distortion to the extracted messages, does not produce  drastic improvement 
of the security. 

Now let us consider an example of more complicated message signal. 
In our simulation, we construct a message
\begin{equation}
I(t)=\frac{A}{m}\sum_{i=1}^{m} B_i\sin (2\pi t/T_i),  
\end{equation} 
where $B_i$ and $T_i$ are random numbers uniform  on $(0, 1)$ and $(50,
500)$ respectively.  
 
Fig. 4 and 5 are results of message extraction for an realization of such
a complicated message with $A=0.01$ and $m=100$. Again it is seen that 
the quality of the recovered message deteriorates more when  masking 
Scheme II is employed. 
However, a comparison  between the power spectra of the original  and 
extracted messages has shown that unmasking is successful for both masking 
schemes.  

We should point out that the  identification of the time-delay $\tau_0$ and the 
quality of the recoved message is not sensitive to the choice of 
$N$, $h$ and $\epsilon$. In general, $\epsilon$ should be small enough to
apply local linear approximation, but large enough to average out
the fluctuations induced by the message.  
As a result, if the amplitude
of the message is too large, the quality of the recovered message can be
quite poor, and the message extraction becomes more difficult. However, the chaotic
signals may not provide enough masking for messages with quite large amplitude.

For the M-G time-delay system  studied  above,  
the frequency of the message should be
rather low, because the power spectrum of the chaotic signal is very low 
at high frequency and is not enough to mask messages with high frequency.
A low frequency of the message means 
$|\dot{I}|\ll |I|$, which is an advantage  for  a third party  to recover 
the message with high quality.

In summary,  we present a simple method to 
 extract messages masked by chaotic signal of a time-delay 
system, which has a very high dimensionality and  many positive Lyapunov exponents.
Using a special embedding space, the infinite dimensional phase space
of the time-delay system is projected to a three-dimensional space,
independent of the actual dimension and the number of positive 
Lyapunov exponents  of the chaotic attractor. The time-delay of the
system is correctly identified even in the presence of the message, which
enables  us to extract the message successfully using a simple local
reconstruction of the time-delay system in the three-dimensional space.

We come to the conclusion based on our analysis that 
communication using time-delay system is not as  secure as 
intuitively expected. In general, the security of chaos communication
may be spoiled if any reconstruction of the dynamics of the system is
possible in some appropriate space, even for very high dimensional dynamics.

\bigskip
{\bf Acknowledgements:}

This work was supported in part by research grant No. RP960689 at the National
University of Singapore.  Zhou is supported by NSTB.

\newpage
{\large \bf Figure Captions}
\begin{description}
\item Fig. 1 As a measure of the smoothness, the average fitting error $E$, as a
function of the embedding delay $\tau$, has a pronounced minimum at the value of
the time-delay of the system. 

\item Fig. 2 Illustration of message extraction for a sine wave  message masked by
Scheme I. (a) The original message signal $kI$, (b) the extracted
message, and (c) the power spectrum  of the extracted message.  

\item Fig. 3 Illustration of message extraction for the  sine wave  message masked
by Scheme II. (a) A time series of the transmitted signal $s$, (b) 
$\Delta s=\dot{s}-\hat{\dot{x}}^i$, (c) the extracted message, and 
(d) the power spectra of the extracted message.

\item Fig. 4 Illustration of message extraction for a complicated message. 
(a) The original message,  (b) the extracted message for the masking Scheme I,  
and (c)the extracted message for the masking Scheme II.

\item Fig. 5 The power spectra of the original message and the extracted messages in
Fig. 4. 

\end{description}

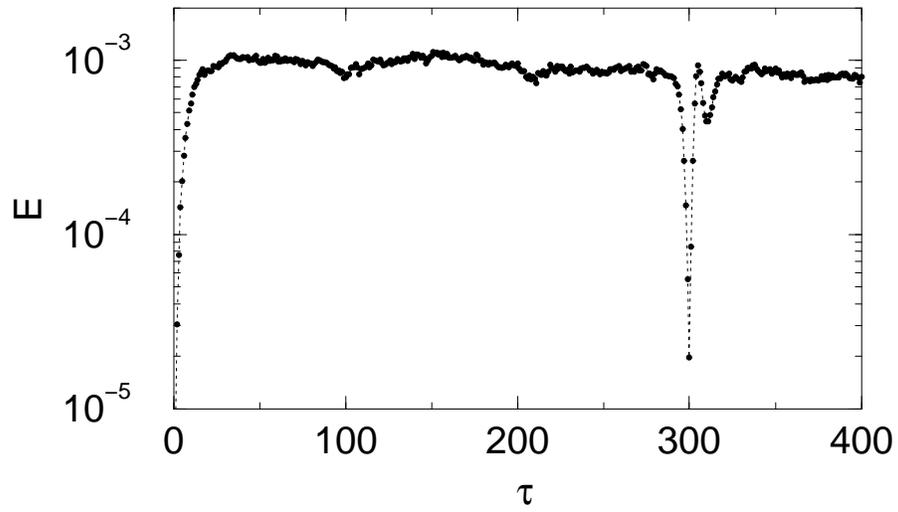
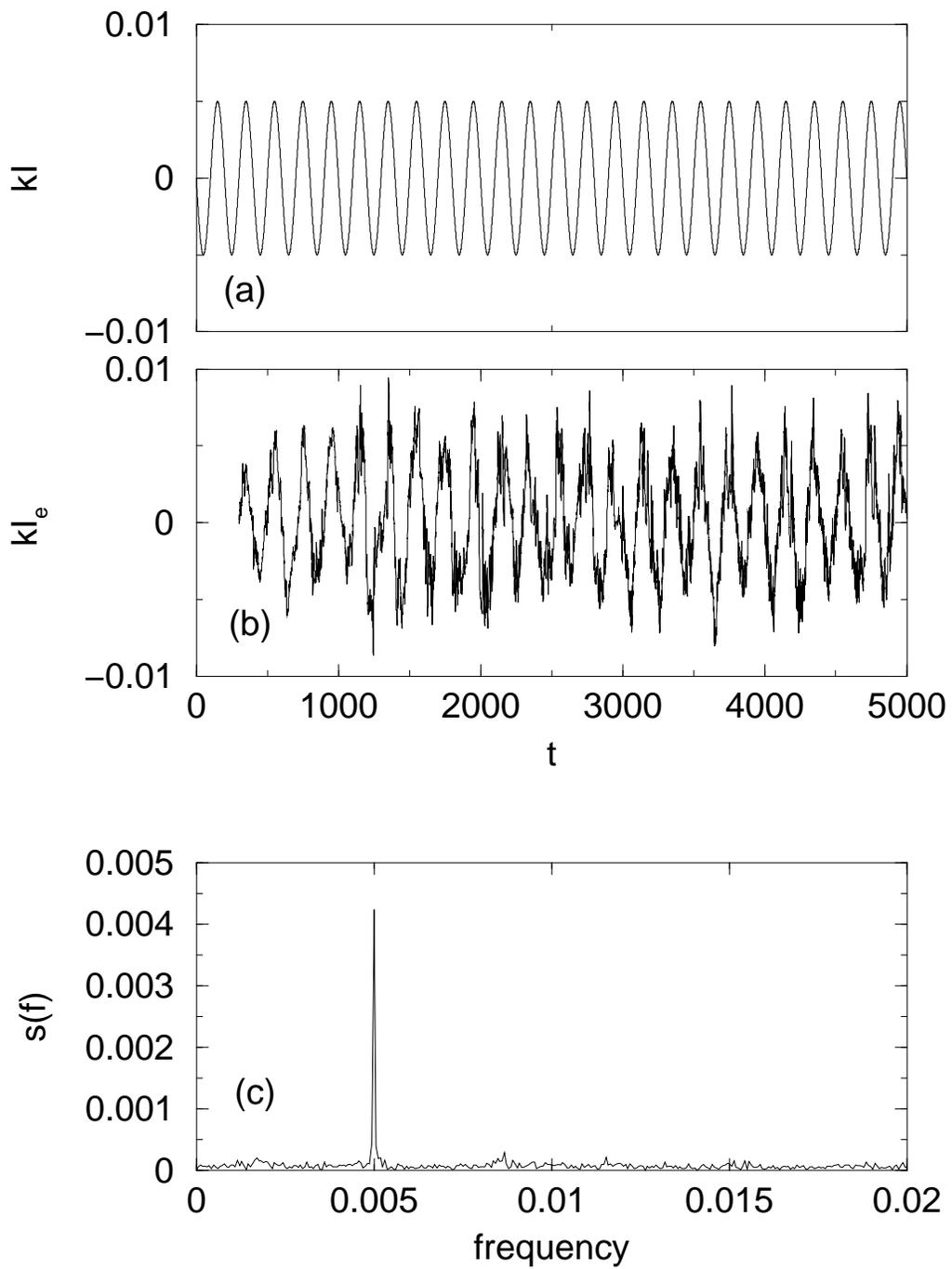
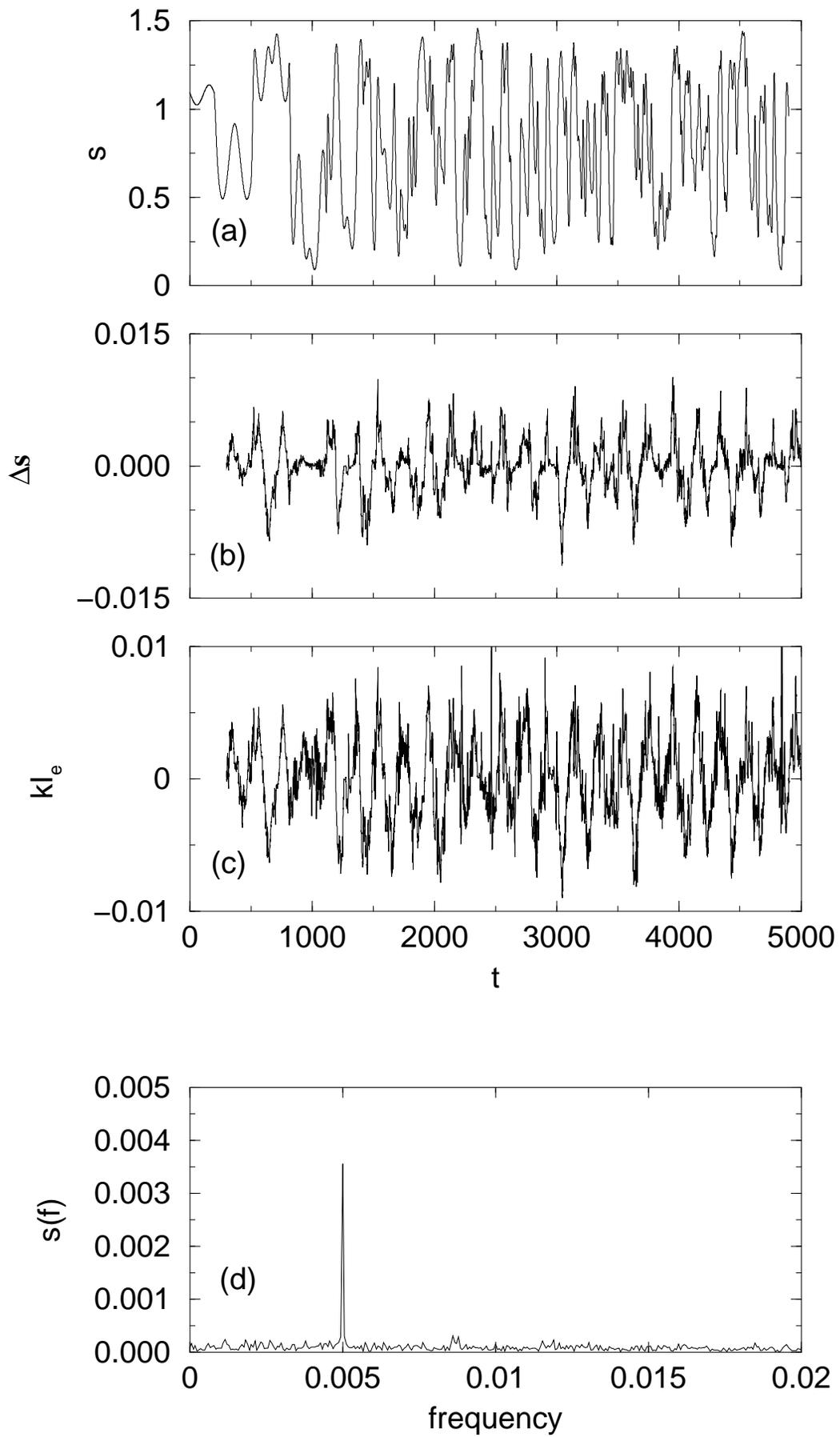
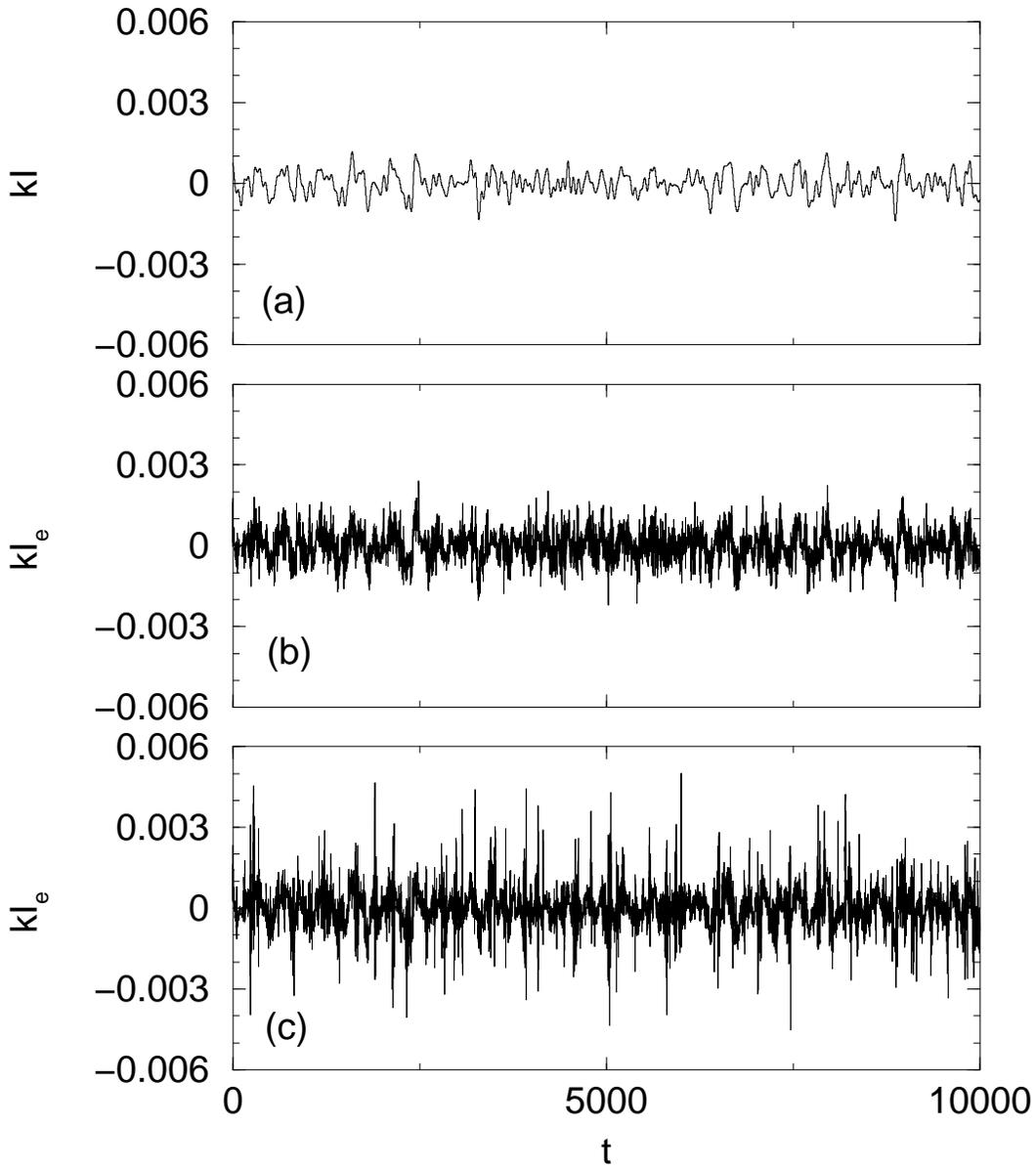
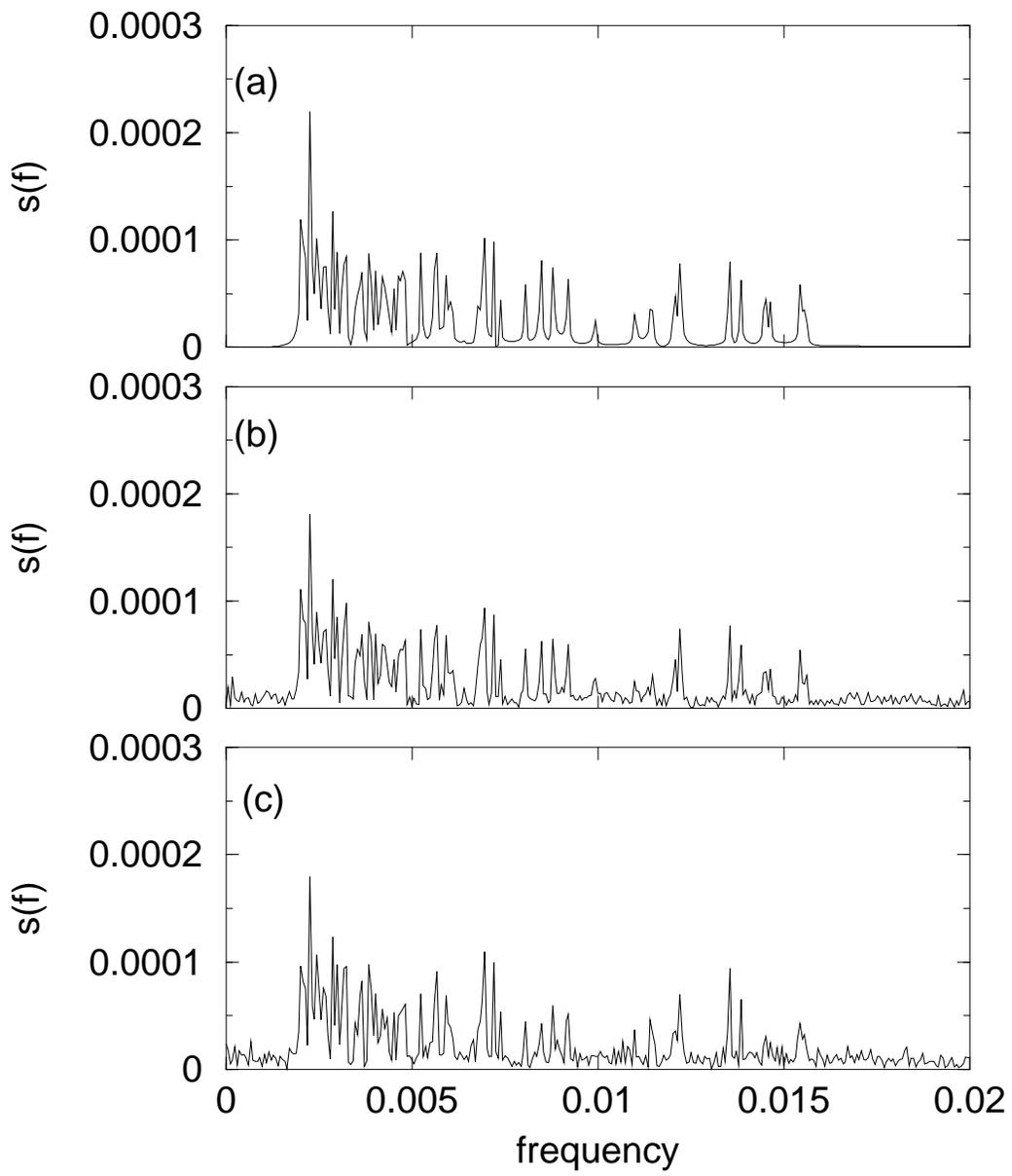
\end{document}